\documentclass[oneside]{article}
\usepackage[T1]{fontenc}
\usepackage{authblk}
\usepackage{amsmath}
\usepackage{graphicx}
\usepackage[english]{babel}
\usepackage{lmodern}
\usepackage[T1]{fontenc}
\usepackage[latin9]{inputenc}

\usepackage{graphicx}
\usepackage[unicode=true,
 bookmarks=false,
breaklinks=false,pdfborder={0 0 1},backref=false,colorlinks=false]
 {hyperref}
\usepackage[a4paper, total={210mm,297mm},margin=2.0cm]{geometry}

\usepackage{datetime}
\newdate{date}{20}{4}{2018}

\title{Lattice Propagators and Haldane-Wu Fractional Statistics} 

\author[1,2,3]{Sauro Succi \thanks{Electronic address: \texttt{s.succi@iac.cnr.it}; Corresponding author}}
\author[2]{Marco Lauricella}

\affil[1]{Center for Life Nano Science @Sapienza, Istituto Italiano di Tecnologia - 295 Viale Regina Elena, I-00161 Rome, Italy}
\affil[2]{Istituto per le Applicazioni del Calcolo CNR, Via dei Taurini 19, 00185 Rome, Italy}
\affil[3]{Institute for Applied Computational Science, Harvard John A. Paulson School of Engineering and Applied Sciences - Cambridge, MA 02138, USA}

\date{\displaydate{date}}

\begin{document}

\maketitle
 
\begin{abstract}
We point out a formal analogy between lattice kinetic propagators and 
Haldane-Wu fractional statistics. 
The analogy could be used to compute the partition function of fractional
quantum systems by solving a corresponding lattice kinetic equation 
for classical dissipative flowing systems.
\end{abstract}

\vspace{1cm}

The analogy between the partition function of a statistical system
at equilibrium and the time-propagator of evolution equations has
been noted since long, and makes the basis for a number of
theoretical insights as well as powerful numerical methods, especially those
based on the path integral formalism \cite{GP}. 

Consider a classical evolution problem governed by the first-order Liouville equation:
\begin{equation}
\partial_t \psi(x;t) = -\mathcal{L} \psi(x;t)
\label{eq:1}
\end{equation}
where $x$ is a phase-space coordinate and $\psi$ is a real scalar
describing the dynamic state of the system.
The formal solution is given by
\begin{equation}
\psi(x;t) = e^{-\mathcal{L}t} \psi(x;0) , 
\label{eq:2}
\end{equation}
where $\psi(x;0)$ is the initial datum and the Liouville operator 
$\mathcal{L}$ is assumed to be non-negative definite on stability accounts.

The stationary regime associated with the Liouville equation, assuming it exists, is described
by a time-independent solution $\psi(x)$, obeying the relation:
\begin{equation}
\{H,\psi(x)\}=0 \; ,
\end{equation}
where $\{ \cdot \}$ denotes the Poisson bracket, and
$\mathcal{H}$ is the Hamiltionian of the system.
Note that, by definition, the Liouville operator is given by  the Poisson bracket
of the Hamiltonian, i.e. $L=\{ H,  \cdot \}$.

A simple additive ansatz reads $\psi(x)=Z^{-1} e^{-\beta \mathcal{H}}$,
with $Z$ a normalization factor and $\beta$ a positive constant.

The formal solution in Eq. \ref{eq:2} also invites an interesting analogy with the partition function of the classical
statistical system $Z(\beta)= <e^{-\beta \mathcal{H}}>$, where $\beta=1/k_BT$ is the inverse
temperature, and brackets denote integration over the microscopic degrees of freedom of the Hamiltonian..
The analogy is highlighted by computing the auto-correlation function
$C(t,\tau) = \int \psi(x;t+\tau) \psi(x;t) dx  = C(t) <e^{-\mathcal{L}\tau}>$, where brackets 
denote averaging over the distribution $\rho(x,t)=\psi^2(x;t)/\int \psi^2(x;t)dx$ and
$C(t) = \int \psi^2(x;t)dx$ \cite{allen1980calculation}.
The natural identification is $\beta = \tau$ and $\mathcal{H}=\mathcal{L}$, indicating that
long-time behaviour of the autocorrelation associates with 
low-temperature partition functions \cite{JPB}.

Note that the weight factor $e^{-\beta \mathcal{H}}$ also includes quantum 
mechanical systems in imaginary time, with the identification 
$k_BT=\hbar/\tau$ and $\mathcal{L} = i \mathcal{H}/\hbar$ (imaginary time).

As mentioned above, this is a direct consequence of the additivity of continuum time, which 
singles out the exponential as the only functional form supporting the group properties
$\mathcal{P}_{a} \mathcal{P}_{b} = \mathcal{P}_{b} \mathcal{P}_{a} = 
\mathcal{P}_{a+b}$ and $\mathcal{P}_{a} \mathcal{P}_{-a} = 1$, 
$\mathcal{P}_t \equiv e^{-\mathcal{L}t}$ being the continuum time-propagator.

In the sequel, we shall show that even though discrete space-time propagators 
are generally non-additive, they still support statistical analogies, encompassing not only
the three standard statistics, Maxwell-Boltzmann, Fermi-Dirac and Bose-Einsten,
but also fractional statistics associated to exotic anyons \cite{lerda2008anyons}.
As we shall see, the basic parameter defining the given fractional
statistics is the relaxation time of the Liouvillean operator
(with dissipative systems in mind, most notably classical fluids). 


A large variety of non-linear field theories, in particular hydrodynamics, can 
be encoded within a lattice Boltzmann equation (LBE) of the form \cite{LB1,benzi1992lattice}:
\begin{equation}
\label{LBE}
f_i(\vec{r}+\vec{c}_i \Delta t,t+\Delta t) -f_i(\vec{r};t) = \frac{\Delta t}{\tau} (f_i^{eq}-f_i)(\vec{r};t)
\end{equation}
where $f_i(\vec{r};t)$ represents the probability of finding a particle at the lattice 
position $\vec{r}$ at time $t$ with discrete velocity $\vec{c}_i$, $i=0,b$.

The LHS represents the free-streaming from site $\vec{r}$ to site
$\vec{r}_i = \vec{r}+\vec{c}_i \Delta t$ in a time lapse $\Delta t$, while the RHS
represents the effects of collisions in the form of a relaxation to the
local equilibrium $f_i^{eq}$ on a time scale $\tau$.

We wish to reiterate that the above formalism is based on the assumption that
the system exhibits sufficient universality to i) Admit one-body local equilibria, whose
space-time dependence is carried solely by slow hydrodynamic modes, 
ii) Relax to such local equilibria according to a constant
time scale, (easily generalised to the case of a multi-relaxation matrix).

While such an assumption might set a restriction on the class of admissible many-body Liouvilleans, it is nonetheless
sufficiently general to embrace an amazingly broad class of linear and non-linear non-equilibrium transport
phenomena, characterised by a weak-departure from local equilibrium,  as it is typical of the hydrodynamic regime
\cite{EPL2038}.

A similar statement holds for quantum mechanics in imaginary time, with 
local equilibria replaced by the ground state of the quantum system, an well-known analogy
that lies at the heart of the Diffusion Monte Carlo method \cite{DMC}.
Even more interestingly, the lattice kinetic formalism, eq. (\ref{LBE}), also extends to 
the {\it real-time} evolution of relativistic and non-relativistic quantum systems, 
as detailed in \cite{FILLI}. This speaks clearly for the generality of the mathematical 
framework associated with the lattice kinetic equation (\ref{LBE}).

For a classical system, the local equilibrium is given by a Maxwell-Boltzmann
distribution in velocity space:
\begin{equation}
f_i^{eq}  =  \rho p(v_i)
\end{equation}
where $p(v_i)$ is a polynomial truncation of the 
gaussian distribution $e^{-v_i^2/2}$ (for non-quantum fluids), 
$v_i=|(\vec{c}_i-\vec{u})|/V_{th}$ and $V_{th}=\sqrt{k_BT/m}$ is the thermal speed.

Note that $f^{eq}$ depends on space and time only through the slow-conserved hydrodynamic
modes, the fluid density, momentum and temperature, namely:
$\rho(\vec{r};t)                              = m \sum_i f_i(\vec{r};t)$, 
$\rho(\vec{r};t) \vec{u}(\vec{r};t)   =  m \sum_i \vec{c}_i f_i(\vec{r};t)$, 
$\rho(\vec{r};t) V^2_{th}(\vec{r};t)= m \sum_i f_i(\vec{r};t) (\vec{c}_i-\vec{u})^2/2$. 

Once the set of discrete velocities is chosen so as to ensure sufficient symmetry to
recover rotational invariance (isotropy), the above LBE can be shown to reproduce 
macroscopic hydrodynamics in the large-scale long-time limit
of small Knudsen numbers, i.e. molecular mean free-path much smaller than the 
macroscale in space and time.
Under such conditions, a second order expansion in the Mach number $u/V_{th}$ of the local
equilibria is sufficient to secure compliance with the isothermal Navier-Stokes equations, so
that the local equilibrium is a local, quadratic mapping of the actual distribution function. 

Symbolically, $$f_i^{eq}=M(f_i),$$ where $M$ is the local quadratic map.  

The kinematic viscosity of the lattice fluid is given by:
\begin{equation}
\label{VISCOP}
\nu = C_s^2 (\tau-\Delta t/2)
\end{equation}
where $C_s$ is the sound speed.
Note that, owing to the lattice discreteness, the viscosity of the
LB fluid vanishes at a finite value of the relaxation time, namely 
$\tau = \Delta t/2$. 

The LB fluid viscosity is often represented in lattice units as follows:
\begin{equation}
\label{VISCOL}
\nu = c_s^2 \frac{\Delta x^2}{\Delta t} (1/\omega-1/2)
\end{equation}
where $\Delta x$ is the lattice spacing and we have set $\omega \equiv \Delta t/\tau$.
In the above $c_s^2$ is the sound speed in lattice units, typically $1/3$ for most LB models.
By stability requirements, the LBE scheme operates in the range $0 \le \omega \le 2$, 
corresponding to a finite, non-negative viscosity.

Three distinguished limits are immediately apparent: 
\begin{enumerate}
\item $\omega \to 0$, ($\nu \to \infty$);
\item $\omega \to 1$, ($\nu \to \nu_l$); 
\item $\omega \to 2$, ($\nu \to 0$).
\end{enumerate}
where $\nu_l \equiv c_s^2 \Delta x^2/2 \Delta t$ denotes the 
characteristic lattice viscosity. 

We hasten to note that, from a hydrodynamic viewpoint, the divergence of the expression
(\ref{VISCOL}) for $\omega \to 0$ is purely formal, since in the continuum limit 
$\Delta t/\tau \to 0$, the correct value of the fluid viscosity 
is $\nu_p = C_s^2 \tau$, according to the expression (\ref{VISCOP}).

This divergence signals the fact that in a regime where the relaxation 
time $\tau$ is much larger than the lattice spacing $\Delta t$, the LB scheme
is not supposed to recover any hydrodynamic behavior, simply because collisions
are too infrequent to promote collective motion.
However, since the present work is by no means restrained to LB as
a hydrodynamic solver, the regime of virtually infinite viscosity 
remains relevant to our discussion.

The interpretation of these distinguished limits becomes particularly
informative by recasting the LBE in the following compact form:

\begin{equation}
\hat f_i = f'_i 
\end{equation}
where 
$\hat f_i \equiv f_i(\vec{r}+\vec{c}_i \Delta t,t+\Delta t)$ 
is the spacetime shifted (post-streaming) distribution and 
$f'_i \equiv (1-\omega) f_i + \omega f_i^{eq}$ denotes the post-collisional distribution.

Case 1 corresponds to no interaction, $f'=f$, no relaxation to local equilibrium,
no fluid, but just a collection of independent particles.
Case 2 corresponds to $f'=f^{eq}$, i.e. the post-collisional state is 
set to the equilibrium, the quickest path to equilibrium. 
Case 3 corresponds to maximum change, $f'=f-2f^{neq}$, the strongest interaction,
leading to zero diffusivity.

The dictionary versus standard numerical analysis is as follows:
\begin{enumerate}
\item Under-Relaxation (Weak-Interaction)

$ 0 \le \omega < 1$;
\item No-Relaxation (Quick-Interaction) 

$ \omega=1$;
\item Over-Relaxation (Strong-Interaction)

$1 < \omega \le 2$;
\end{enumerate}

It is instructive to recast the lattice viscosity in fully symmetric form as
\begin{equation}
\nu /\nu_l = \; \frac{1-\epsilon}{1+\epsilon}
\end{equation}
where $\epsilon \equiv \omega -1$.
From the above, it is seen that the lattice viscosity is dual around $\epsilon=0$, i.e.
$\nu(-\epsilon) = 1/\nu(\epsilon)$, a property which evokes the reversibility of group
operators.  

\begin{figure}[!ht]
\begin{centering}
\includegraphics[width=0.6\columnwidth]{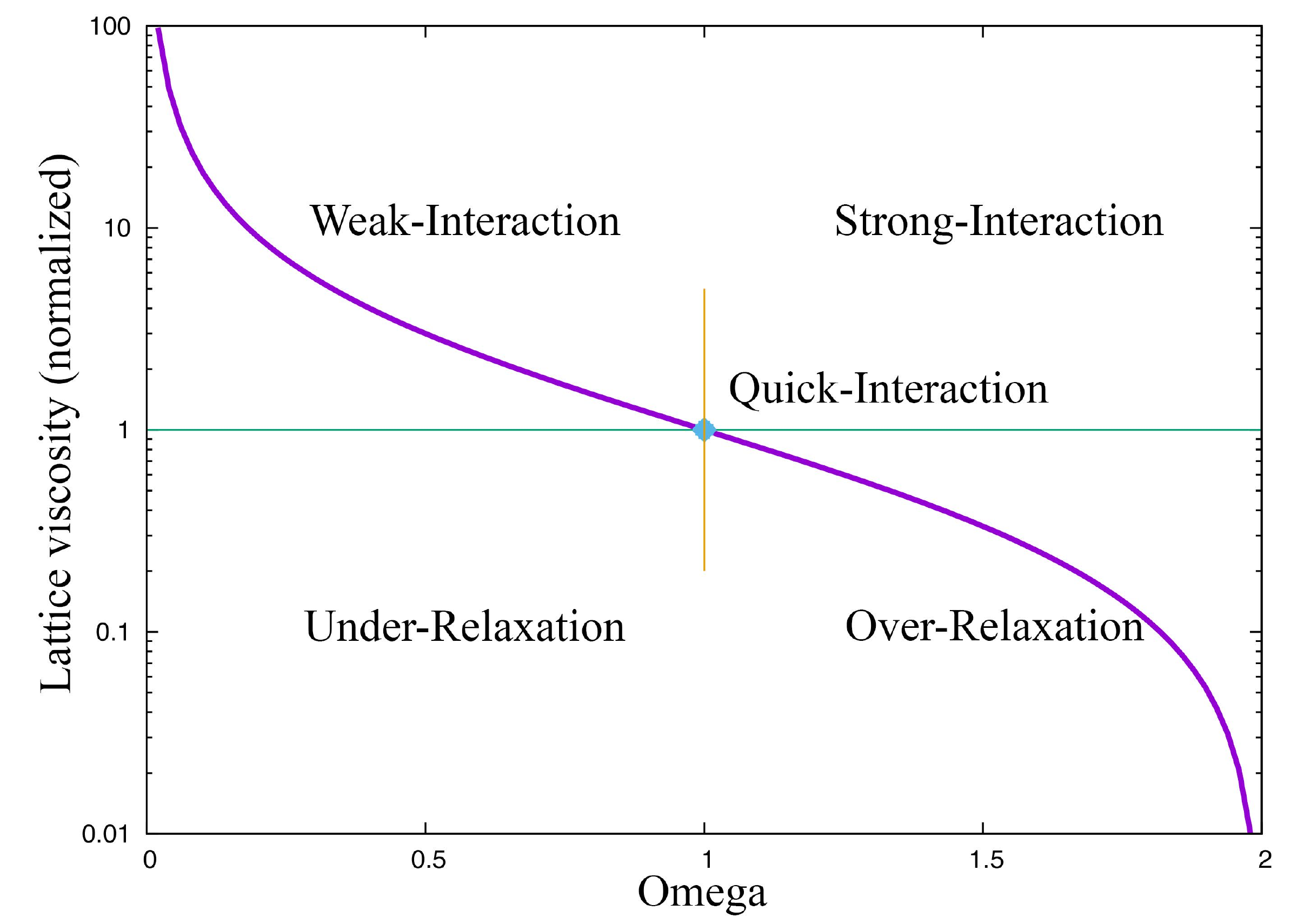}
\par\end{centering}
\caption{The normalised lattice fluid viscosity, $\nu/\nu_l$ as a function of the relaxation parameter $\omega$.
The dual structure around the quick-interaction value $\omega=1$ is apparent.
For a geometrical interpretation in terms of entropy minimization see \cite{BRUCE}.
}
\end{figure}
As we shall show shortly, these limiting cases map onto the three
main statistics, Bose, Maxwell and Fermi, respectively.

Before going into these matters, in passing we note
a suitable choice of the local equilibria gives rise to different
types of non-linear PDE's, not just fluids \cite{LB2038}, including quantum ones \cite{FILLI}.
In this respect, the LBE can be regarded as a functional generator, lifting
nonlinear field theory to a higher-dimensional kinetic space, whereby
field theory emerges as a large scale, long-time, limit projection
of the higher dimensional lattice kinetic theory.
The extra dimensions are exposed by the non-equilibrium component of the distribution
which depends not only on the conserved modes but also on their
space-time derivatives to all orders (non-local, non-universal terms), associated 
with increasing powers of the Knudsen number.


To elicit the role of the Knudsen number in probing extra-dimensions of kinetic space, it
proves expedient to recast eq. (\ref{LBE}) in the following compact integral form:
\begin{equation}
\label{GLBE}
f_i(\vec{r},t) = G(k_i,\omega) \; f_i^{eq}(\vec{r};t)
\end{equation}
where the "Green function" reads as
\begin{equation}
\label{GREEN}
G(k_i,\omega)  =  \frac{1}{1+\frac{\tau}{\Delta t} (e^{\Delta t \partial_i}-1)}
= \frac{\omega}{e^{k_i} -1 + \omega}
\end{equation}
In the above, $k_i \equiv \Delta t \partial_i$, is the lattice Knudsen operator along 
the i-the direction associated with the directional covariant derivative
$
\partial_i = \partial_t + \vec{c}_i \cdot \nabla.
$
The continuum limit $\Delta t/\tau \to 0$, yields 
$G(k_i,\omega) \to 1/(1+\tau \partial_i)$, corresponding 
to the continuum time semi-discrete Boltzmann-BGK equation:
$$
(1+ \tau \partial_i) f_i = f_i^{eq}
$$
Interestingly, the lattice version can be written exactly in the same form
by simply replacing $\partial$ with its lattice-deformed 
analogue $\partial_{\Delta} \equiv (e^{\Delta t \partial}-1)/\Delta t$. 
This offers an elegant example of lattice regularization-renormalization, in the sense
that LB can be regarded as a coarse-grained Boltzmann equation in discrete phase-space
whereby the covariant derivative $\partial$ is replaced by its lattice-deformed 
counterpart. The result is the appearance of a negative contribution to 
the viscosity, as expressed by the factor $-1/2$ at the right hand side of eq. (\ref{VISCOL}).

The Green function maps the local equilibrium into the actual distribution by 
resumming an infinite series in the Knudsen number, each term incorporating 
higher orders of inhomogeneity and non-locality \cite{KARGOR}.

Indeed, the Green function exhibits an interesting series expansion
$
G(k,\omega)  = \sum_{n=0}^{\infty} g_n(\omega) k^n
$
where the polynomial coefficients $g_n(\omega)$ are related 
to the Bernouilli polynomials through the relation 
$z/(e^z-1)=e^{-zt}\sum_{n=0}^\infty \frac{B_n(t)}{n!} z^n$.
The first three are $g_0=1$, $g_1=-\theta$, $g_2=2\theta^2-\theta$, 
and so on \cite{NOBLE}, where we have set $\theta \equiv 1/\omega = \tau/\Delta t$.
In other words, the Green function (\ref{GREEN}) is the generating function 
of the $g_n(\omega)$ polynomials, each polynomial carrying the dependence 
on the Knudsen number at the corresponding order. 
This is a lattice version of the continuum Green function $(1+\tau \partial)^{-1}$, whose
generating polynomials are $(-\tau)^n$.

It is important to note that the above Green-function differs from 
a standard time-propagator in that, instead of
propagating the state from time $t$ to time $t+\Delta t$, it generates the actual distribution
at time $t+\Delta t$ starting from the local equilibrium at the time $t$.
In other words, it bootstraps the non-equilibrium component via an all-term
application of the spacetime gradient (streaming operator) to the equilibrium 
distribution at the previous time-step.
\begin{figure}[!ht]
\begin{centering}
\includegraphics[width=0.6\columnwidth]{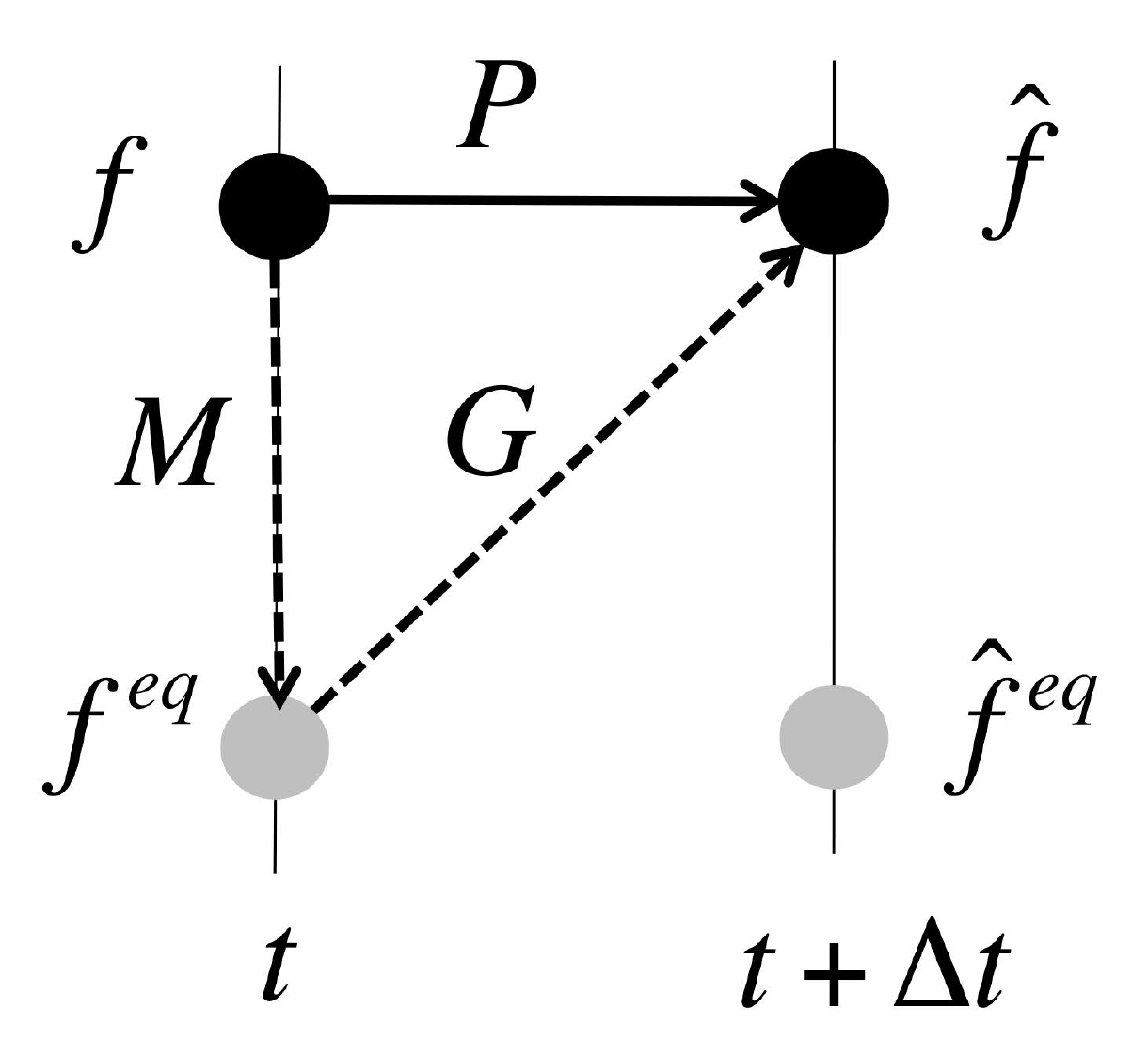}
\par\end{centering}
\caption{The time-propagator scheme of the LB method. 
The actual distribution $f$ at time $t$ forms the corresponding 
equilibrium $f^{eq}$ via the local nonlinear mapping $M$ given by eq. (3). 
The local equilibrium at time $t$ is then propagated to the actual distribution
$\hat f$ at time $t+\Delta t$ via the Green function $G$. 
Formally, the time-propagator taking $f$ from time $t$ to $\hat f$ at
time $t+\Delta t$, is given by the product $P=GM$
This zig-zag $GM$ scheme repeats cyclically in time to cover 
the entire time-span of the simulation. 
The distinguished cases $\Delta t \to 0,\tau,2 \tau$, recover 
the Bose, Maxwell and Fermi statistics, respectively. 
}
\end{figure}

It is readily observed that the three distinguished cases $\omega=0,1,2$ correspond
to the three major statistics, namely:

\begin{enumerate}
\item Bose: $\omega=0,\epsilon=-1$, $G_0 = \omega /(e^k-1)$
\item Maxwell: $\omega=1$, $\epsilon=0$, $G_1 = e^{-k}$
\item Fermi: $\omega=2,\epsilon=+1$, $G_2 = 2 /(e^k+1)$
\end{enumerate}

To be noted that the Bose case corresponds to continuum 
time, $\Delta t /\tau \to 0$, in which case both numerator and numerators 
of $G_0$ vanish, recovering the continuum Green function $1/(1+ \tau \partial)$
as a proper limit, exposing the physical Knudsen number $\tau \partial$ instead of
the lattice one, $\Delta t \partial$, as it should.  
Interestingly, the generic case $0<\omega<2$ also exhibits an approximate statistical
parallel with the exotic case of anyons, i.e. particles 
with fractional statistics \cite{FW1,FW2}. 


To elucidate this point, let us remind that the Haldane-Wu 
distribution for a fractional excitation of index $a$, obeys the 
relation \cite{HAL,WU}:

\begin{equation}
\label{HW1}
W^a (1+W)^{1-a} = p(z)
\end{equation}
where we have set $p(z) \equiv e^z$ and $z \equiv (E-\mu)/kT$.
In the following, we shall set the chemical potential to $\mu=0$, on the 
assumption that this does not imply any loss of generality.

The probability distribution function (pdf) of anyons with index $0 \le a \le 1$,
($a=0$ Bose, $a=1$ Fermi) is:
\begin{equation}
\label{HW2}
f_a(z) = \frac{1}{W(z)+a} 
\end{equation}

Since $W \ge 0$, $f_a(z) \le 1/a$, as it should be for particles
obeying fractional statistics.
From the physical point of view, the exponent $a$ associates with 
the fractional charge of quasi-particle excitations in one or two spatial dimensions.

The special case of semions ($a=1/2$) can be solved analytically, to deliver
$\label{SEMION}
W_{1/2}(z) = \frac{-1 + \sqrt{1 + 4 p^2(z)}}{2},
$
where the minus sign has been excluded on positiveness grounds
The cases $a=1/3$ and $a=2/3$ can also be solved analytically
via cubic roots. However, the corresponding expressions are rather cumbersome
and shall not be investigated here.
In the low-energy limit,
$
W \ll 1,
$
the expression (\ref{HW1}) delivers $W \sim p^{1/a}$, that is:
$
f_a(z) \sim \frac{1}{p^{1/a}(z)+a}.
$
In the high-energy limit, 
$
W \gg 1,
$
the expression (\ref{HW1}) yields $W \sim p$, and more precisely, to first order 
accuracy in $1/W$, $W=p-1+a$, whence:
$$
f_a(z) \sim \frac{1}{p(z)+2a-1}.
$$
Note that, within this high-energy approximation, the semion pdf with $a=1/2$ matches exactly Maxwell's distribution.

The corresponding pdf's are reported in Fig. 2. 
\begin{figure}[!ht]
\begin{centering}
\includegraphics[width=0.6\columnwidth]{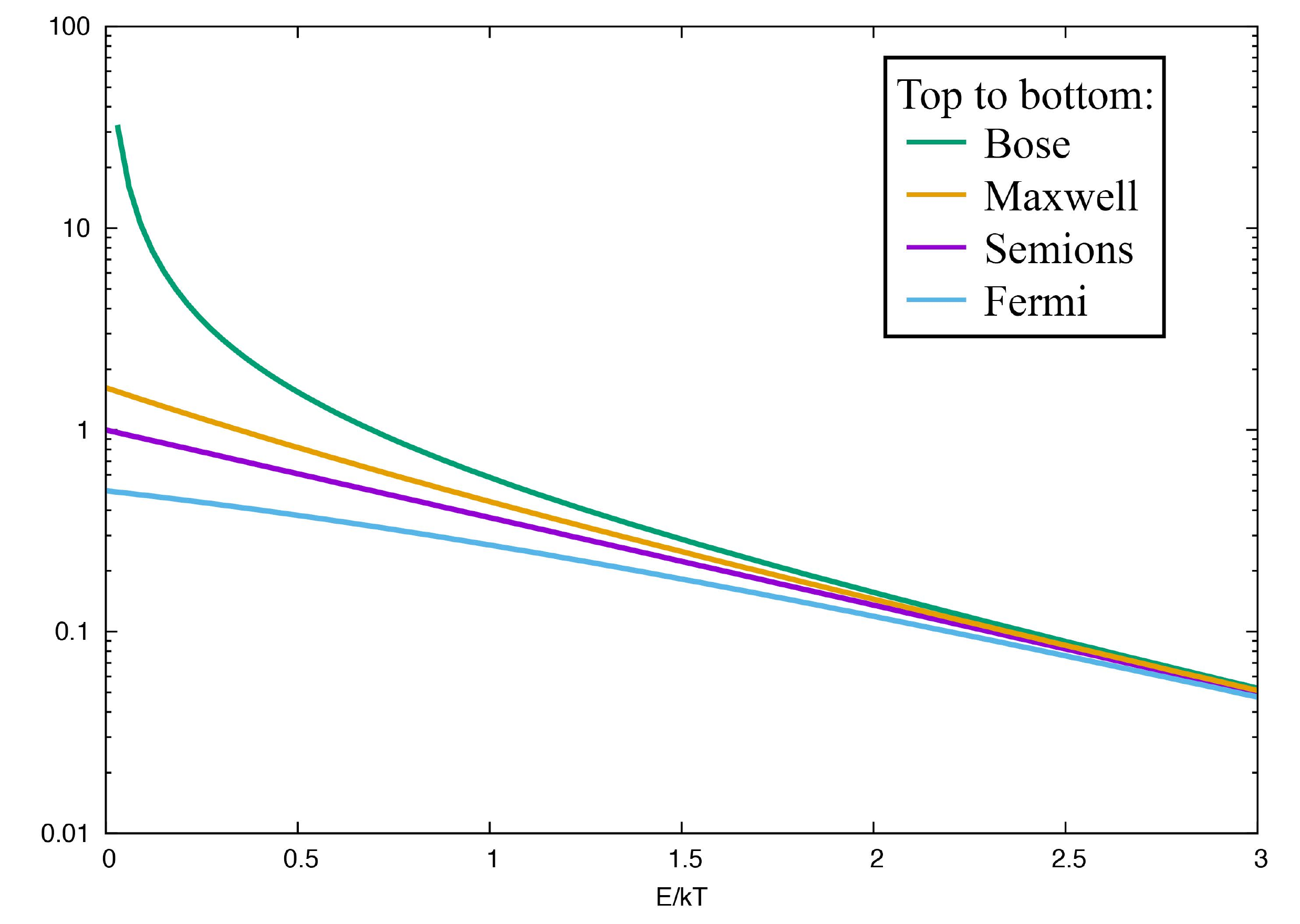}
\par\end{centering}
\caption{The (unnormalized) Bose, Maxwell, Semion and Fermi probability distributions (top to bottom).
The semion distribution remains relatively close to Maxwell also at low energy,
(the lower limit on display is $E/k_BT=0.2$).
}
\end{figure}
Inspection the graph of the function $F_a(w)=w^a (1+w)^{1-a}$ 
for different values $0 \le a \le 1$ shows that $F_a(w)$ transits from 
$w+1$ for $a=0$ to $w$ for $a=1$, without ever significantly departing from a straight 
line $w+k(a)$, with $k(a)$ a fraction of $a$, except in the vicinity of $w=0$ (very low energy limit).

Based on the above observations, we approximate the anyon pdf as:
\begin{equation}
f_a(z) = \frac{1}{p(z)+s(a)} 
\end{equation}
where the shift $s(a)=a+k(a)$ obeys the boundary conditions  $s(0)=-1, s(1)=1$.
As a first order approximation, we take $s(a)=2a-1$, which, as noted before, is 
tantamount to equating the semion statistics to the classical one.

The analogy is now apparent, with the following identifications:
\begin{itemize}
\item $\Delta t \partial \leftrightarrow \beta \mathcal{H}$,
\item $\omega = s(a)+1$
\end{itemize}
The latter simplifies to $\omega = 2a$ in the semion=Maxwell approximation.


The analogy portrayed in this Letter is intriguing and maybe even useful. 

As mentioned earlier on, the analogy between time-propagators
and the partition function permits to compute the latter by calculating the correlation
function of the corresponding evolution Liouville problem, and viceversa.
Likewise, the analogy brought up in this paper could be used to compute the
partition function of anyon systems by solving a corresponding lattice Boltzmann
equation with the corresponding value of the relaxation parameter $\omega$.
The relevant observable to this purpose is the 
Equilibrium-NonEquilibrum (ENE) correlator, namely:
\begin{equation}
G_i(t;\omega,\beta) = \frac{<f_i^{neq},f_i^{eq}>}{<f_i^{eq},f_i^{eq}>} 
\end{equation}
where brackets denote averaging over configuration space.
Here the index $i=0,8$ corresponds to the standard D2Q9 lattice \cite{D2Q9}, with one 
rest particle $\lbrace c_{0x},c_{0y} \rbrace = (0,0)$, 
four particles with speed $1$, $\lbrace c_{ix},c_{iy} \rbrace= 
\lbrace (1,0),(0,1),(-1,0),(0,-1) \rbrace$ and four particles with
speed $\sqrt 2$, namely $ \lbrace c_{ix},c_{iy} \rbrace = \lbrace (1,1),(-1,1),(-1,-1),(1,-1) \rbrace$.


As an example, in Fig.2, we report the time-asymptotic ENE correlator 
of the mainstream ($f_1$), cross-flow ($f_2$)
and diagonal ($f_5$) populations, for the case of a two-dimensional channel 
(Poiseuille) flow,
$u_x(x,y) = U_0 y(1-y)$, $u_y(x,y)=0$, where $x$ is the mainstream 
coordinate and $-1 \le y \le 1$ the normalized cross-flow coordinate. 
The simulation runs on a $128 \times 64$ grid, corresponding to a
resolution $\Delta x/L = 1/64$, $L$ being the cross-flow dimension of the channel.

\begin{figure}[!ht]
\begin{centering}
\includegraphics[width=0.6\columnwidth]{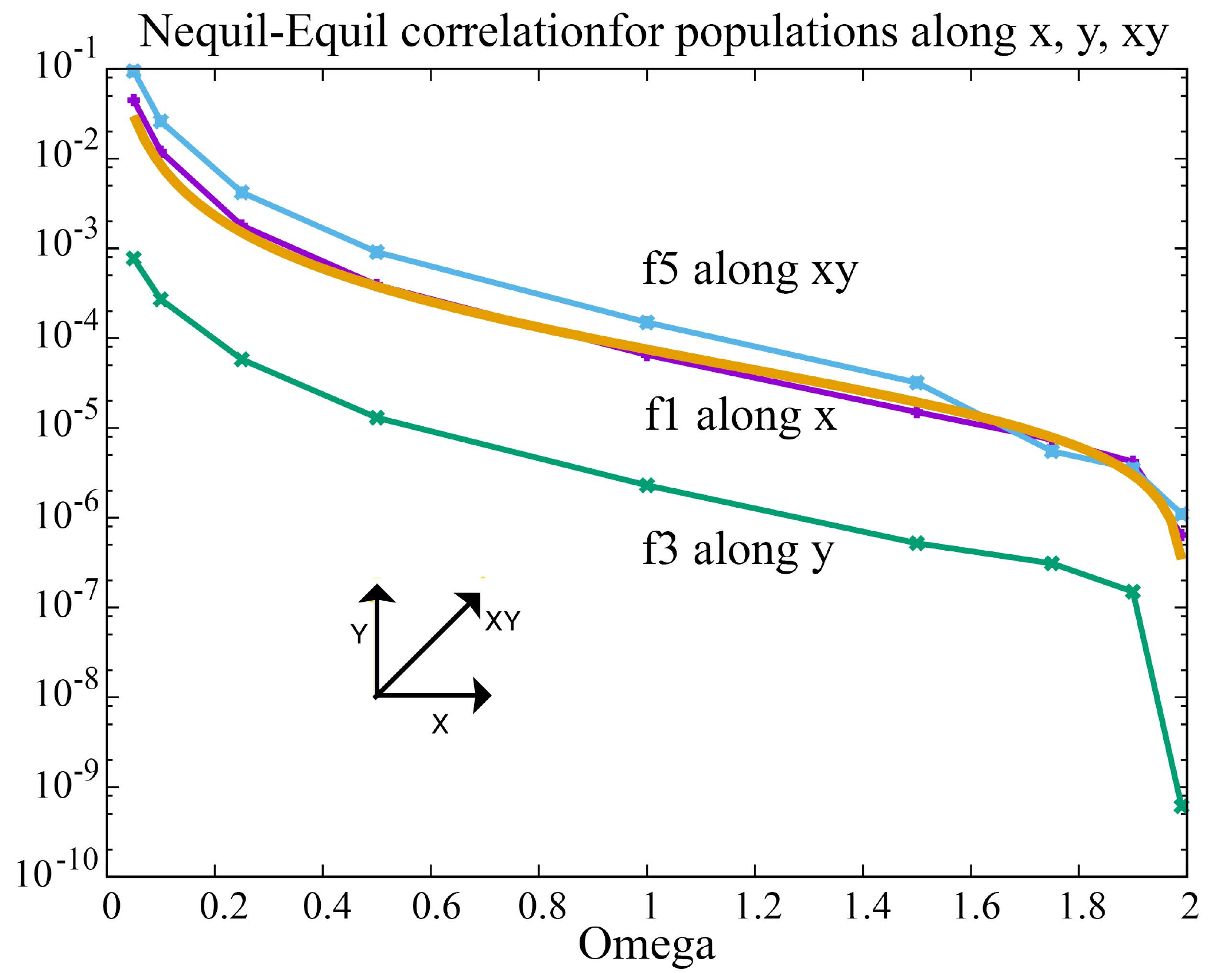}
\par\end{centering}
\caption{The ENE correlator for $x,y,xy$ populations as obtained from a 
Lattice Boltzmann simulation of a two-dimensional channel flow between parallel plates.
The fit to the $xx$-ENE correlator with $A=7.5 \;10^{-5}$, $\alpha = -1.8$ and $\beta=0.9$,
is also reported (solid line).
}
\end{figure}
We note that the ENE correlator is well fitted by 
$G_{fit}(\omega)= A \omega^{\alpha} (2-\omega)^{\beta}$, a distribution 
which correlates with the lattice fluid viscosity $\nu \propto \omega^{-1} (2-\omega)$. 
Although not obvious a-priori, this is not surprising, since the ENE correlator is 
a close relative to the fluid viscosity via the fluctuation-dissipation theorem.
Indeed, similar correlators have been recently shown to play a major role 
in designing a new class of LB models providing enhanced stability 
via compliance with entropic constraints \cite{KBC}.
The departure of the singular exponent around $\omega \to 0$ is also 
understandable, due to the lack of hydrodynamic content of the LB 
scheme in this limit.  
In passing, we note a close resemblance to the ranking distribution of citation records
and other statistical indicators of human performance \cite{GENE}.

It should be noted that the ENE correlator may depend on the specific
flow which is used to compute it. This is plausible, as it reflects the influence 
of the boundary conditions on the corresponding partition function, an effect 
which is hard to capture by analytical methods.

We also remark that, for the sake of simplicity, the LB simulations shown in Fig. 3 
are based on truncated Maxwell-Boltzmann equilibria, but other choices 
can be readily accomodated within the LB formalism.

The analogy between time-propagators and the canonical distribution in classical statistical
mechanics rests on the exponential function as a very special common cornerstone, securing
time and entropy additivity, respectively \cite{TSA}.
In this work, we have shown that in discrete time, where such additivity no longer
holds, the analogy still applies, although in the form of a more general
fractional Haldane-Wu statistics.
The standard Bose-Fermi and Maxwell statistics are recovered as special instances
of the Haldane-Wu distribution for the case of 
strongly-interacting systems (Fermi), weakly-interacting systems (Bose) and
quickly-interacting systems (Maxwell, close to semions).    
The parameter dictating the fractional statistics is the ratio between
lattice timestep (advection timescale in discrete space-time) and 
the collision-relaxation time scale.
This might open intriguing prospects for computing the equilibrium partition function 
of exotic quantum materials with fractional statistics by means 
of lattice kinetic simulations of classical, dissipative flowing systems.

\section*{Acknowledgments}
Valuable discussions with G. Parisi are kindly acknowledged.
The research leading to these results has received funding from the European Research
Council under the Horizon 2020 Programme Grant Agreement n. 739964 ("COPMAT").


\begin{thebibliography}{10}
\expandafter\ifx\csname url\endcsname\relax
  \def\url#1{\texttt{#1}}\fi
\expandafter\ifx\csname urlprefix\endcsname\relax\def\urlprefix{URL }\fi
\expandafter\ifx\csname href\endcsname\relax
  \def\href#1#2{#2} \def\path#1{#1}\fi

\bibitem{GP}
G.~Parisi, Statistical field theory, Addison-Wesley, 1988.

\bibitem{allen1980calculation}
J.~Allen, D.~Diestler, On the calculation of classical time-autocorrelation
  functions: Resolution of the liouville operator, The Journal of Chemical
  Physics 73~(9) (1980) 4597--4612.

\bibitem{JPB}
J.~P. Boon, S.~Yip, Molecular hydrodynamics, Courier Corporation, 1980.

\bibitem{lerda2008anyons}
A.~Lerda, Anyons: Quantum mechanics of particles with fractional statistics,
  Vol.~14, Springer Science \& Business Media, 2008.

\bibitem{LB1}
G.~R. McNamara, G.~Zanetti, Use of the boltzmann equation to simulate
  lattice-gas automata, Physical review letters 61~(20) (1988) 2332.

\bibitem{benzi1992lattice}
R.~Benzi, S.~Succi, M.~Vergassola, The lattice boltzmann equation: theory and
  applications, Physics Reports 222~(3) (1992) 145--197.

\bibitem{EPL2038}
S.~Succi, Lattice boltzmann 2038, EPL (Europhysics Letters) 109~(5) (2015)
  50001.

\bibitem{DMC}
D.~M. Ceperley, B.~Alder, Ground state of the electron gas by a stochastic
  method, Physical Review Letters 45~(7) (1980) 566.

\bibitem{FILLI}
F.~Fillion-Gourdeau, H.~Herrmann, M.~Mendoza, S.~Palpacelli, S.~Succi, Formal
  analogy between the dirac equation in its majorana form and the
  discrete-velocity version of the boltzmann kinetic equation, Physical review
  letters 111~(16) (2013) 160602.

\bibitem{BRUCE}
B.~M. Boghosian, J.~Yepez, P.~V. Coveney, A.~Wager, Entropic lattice boltzmann
  methods, in: Proceedings of the Royal Society of London A: Mathematical,
  Physical and Engineering Sciences, Vol. 457, The Royal Society, 2001, pp.
  717--766.

\bibitem{LB2038}
S.~Succi, Lattice boltzmann 2038, EPL (Europhysics Letters) 109~(5) (2015)
  50001.

\bibitem{KARGOR}
A.~N. Gorban, I.~V. Karlin, Short-wave limit of hydrodynamics: A soluble
  example, Physical review letters 77~(2) (1996) 282.

\bibitem{NOBLE}
D.~J. Holdych, D.~R. Noble, J.~G. Georgiadis, R.~O. Buckius, Truncation error
  analysis of lattice boltzmann methods, Journal of Computational Physics
  193~(2) (2004) 595--619.

\bibitem{FW1}
F.~Wilczek, Magnetic flux, angular momentum, and statistics, Physical Review
  Letters 48~(17) (1982) 1144.

\bibitem{FW2}
F.~Wilczek, Fractional statistics and anyon superconductivity, Vol.~5, World
  Scientific, 1990.

\bibitem{HAL}
F.~D.~M. Haldane, fractional statistics in arbitrary dimensions: A
  generalization of the pauli principle, Physical review letters 67~(8) (1991)
  937.

\bibitem{WU}
Y.-S. Wu, Statistical distribution for generalized ideal gas of
  fractional-statistics particles, Physical review letters 73~(7) (1994) 922.

\bibitem{D2Q9}
Y.~Qian, D.~d'Humi{\`e}res, P.~Lallemand, Lattice bgk models for navier-stokes
  equation, EPL (Europhysics Letters) 17~(6) (1992) 479.

\bibitem{KBC}
I.~Karlin, F.~B{\"o}sch, S.~Chikatamarla, Gibbs' principle for the
  lattice-kinetic theory of fluid dynamics, Physical Review E 90~(3) (2014)
  031302.

\bibitem{GENE}
A.~Petersen, H.~E. Stanley, S.~Succi, Statistical regularities in the
  rank-citation profile of individual scientists 1 (2011) 14004.

\bibitem{TSA}
C.~Tsallis, On the foundations of statistical mechanics, The European Physical
  Journal Special Topics 226~(7) (2017) 1433--1443.

\end{thebibliography}
\end{document}